\documentstyle[aps]{revtex}
\begin{document}
\newcommand{\bm}{\bibitem}
\newcommand{\bgea}{\begin{eqnarray}}
\newcommand{\ndea}{\end{eqnarray}}
\newcommand{\bge}{\begin{equation}}
\newcommand{\nde}{\end{equation}}
\newcommand{\lbl}{\label}
\newcommand{\rf}[1]{(\ref{#1})}
\newcommand{\srf}[1]{\ref{#1}}
\newcommand{\hf}{\frac{1}{2}}
\newcommand{\bb}[1]{{\vec #1}}
\newcommand{\rr}{{\vec r}}
\newcommand{\kvec}{{\vec k}}
\newcommand{\vna}{\bb \nabla}
\newcommand{\lpl}{{\bb \nabla}^2}
\newcommand{\prtt}{\partial_t}
\newcommand{\q}{\bb q}
\newcommand{\hpsi}{\hat{\psi}}
\newcommand{\hj}{\hat{J}}
\newcommand{\pv}[1]{\langle #1\rangle}
\newcommand{\tv}[1]{\overline {#1} }

\title{Phenomenological Theory for Phase Turbulence in Rayleigh-B\'{e}nard  Convection}

\author{Xiao-jun Li $^a$,
	Hao-wen Xi $^b$,
	J. D. Gunton $^{c,}$ 
	\footnote{Corresponding author. Tel: +1-610-758-3959;
	Fax: +1-610-758-xxxx; E-mail address: jdg4@lehigh.edu}}
\address{$^a$ Department of Physics, University of Washington,
Seattle, Washington 98195-1560, USA}
\address{$^b$ Department of Physics and Astronomy,
  Bowling Green State University,
  Bowling Green, Ohio 43403, USA}
\address{$^c$ Department of Physics,
  Lehigh University, Bethlehem, Pennsylvania 18015, USA}
\maketitle

\begin{abstract}
We present a phenomenological theory for phase turbulence (PT)
in Rayleigh-B\'{e}nard  convection, based on the generalized Swift-Hohenberg
model. We apply a Hartree-Fock approximation to PT and conjecture a scaling
form for the structure factor $S(k)$ with respect to
the correlation length $\xi_2$. We hence obtain
{\it analytical} results for the time-averaged convective current $J$ and
the
time-averaged mean square vorticity $\Omega$.
We also define power-law behaviors such  as
$J \sim \epsilon^\mu$, $\Omega \sim \epsilon^\lambda$ and
$\xi_2 \sim \epsilon^{-\nu}$, where $\epsilon$ is the control parameter.
We find from our theory that $\mu = 1$, $\nu \ge 1/2$ and
$\lambda = 2 \mu + \nu$.
These predictions, together with the scaling conjecture
for $S(k)$, are confirmed by our numerical results.
\\PACS:  47.54.+r; 47.20.Lz; 47.20.Bp; 47.27.Te 
\\Keywords:  Rayleigh-B\'{e}nard  convection; phase turbulence;  
	Hartree-Fock approximation; scaling form
\end{abstract}

\section{Introduction} \lbl{int}
Rayleigh-B\'enard convection (RBC) has long been a
paradigm in the study of
pattern formation \cite{cr_ho_93,ah_95}. This system consists
of a thin horizontal layer of fluid heated from below.
There are three dimensionless parameters to describe the system
\cite{ch_81}.
The Rayleigh number $R \equiv g \alpha d^3 \Delta T/\kappa \nu$ is the
control parameter,
in which $g$ is the gravitational acceleration,
$d$ the layer thickness, $\Delta T$ the temperature gradient
across the layer,
$\alpha$ the thermal expansion coefficient, $\kappa$ the thermal diffusivity
and $\nu$ the kinematic viscosity.
Under the Boussinesq approximation, only the density of the
fluid is temperature dependent.
Then the Prandtl number $\sigma \equiv \nu/\kappa$ is all one needs to
specify the fluid properties.
The third parameter is the aspect ratio
$\Gamma \equiv L/d$ where $L$ is the horizontal size of the system.
When the Rayleigh number $R$  surpasses a critical value $R_c$,
the fluid bifurcates from a static conductive state to a convective state,
in which the velocity profile $\bb u = ({\vec u}_\perp, u_z)$
and the temperature-deviation profile $\theta$
form certain self-organized patterns.
Those patterns
also depend on the boundary conditions (b.c.) at the
horizontal surfaces of the container.
The two most studied  b.c. in the literature are the
{\it rigid-rigid}  b.c., under which  the fluid cannot slip,
and the {\it free-free}  b.c.,
under which the fluid does not experience any stress.

The patterns and the corresponding stability domain in RBC have
been studied extensively
in the classical work of Busse and his coauthors \cite{sc_lo_65,bu_bo_84}
in the $(R, \sigma, k)$ space with $k$ the wavenumber.
For free-free boundaries, Zippelius and Siggia
\cite{zi_si_82} and Busse and Bolton \cite{bu_bo_84} found that
the parallel roll state is unstable against the skewed-varicose
instability immediately above onset if $\sigma < 0.543$. Busse
{\it et al.} \cite{bu_89} further investigated the dynamics
involved and conjectured a direct transition from conduction to
{\it spatiotemporal chaos} (STC). This spatiotemporally chaotic  state is called {\it
phase turbulence} (PT). Recently we reported a large scale
($\Gamma = 60$) numerical simulation of the three dimensional
hydrodynamical equations for $\sigma = 0.5$ under the free-free
b.c. \cite{xi_li_97,xi_li_00}. From that simulation, we confirmed the
direct transition to PT above onset and studied various
properties of it. The patterns we found have very complicated
spatial and temporal dependences.

In this paper we develop a theory for PT based on the generalized
Swift-Hohenberg (GSH) model of RBC \cite{sw_ho_77,cr_80,si_zi_81}.
This model is derived from the three-dimensional hydrodynamic
equations, but is much simpler to study both numerically and
analytically and is widely used in theoretical studies.  
Numerical solutions of this model or its modified
versions have not only reproduced most patterns observed in
experiments but also resembled experimental results relatively
well \cite{cr_ho_93,xi_vi_92,xi_gu_93,xi_gu_95}. But there are
some shortcomings in the model \cite{gr_cr_85}: The stability
boundary of the model does not coincide with that of
hydrodynamics for both rigid-rigid and free-free boundaries; it
induces an unphysical, short-ranged cross roll instability.
Even so, owing
to its simplicity and its qualitative resemblance to real
systems, this model is very valuable in studying RBC.

There are two coupled equations 
in two-dimentional space $\rr = (x,y)$ in the GSH model  
for two order parameters
$\psi(\rr,t)$ and $\omega_z(\rr,t)$, which are, respectively, 
related to the convective
current and the vertical vorticity \cite{li00}.
In this paper we present our {\it analytical} calculations
of the
time-averaged convective current $J = A^{-1} \int d
\bb r\,\tv{\psi^2(\bb r,t)}$ and the time-averaged mean square
vorticity $\Omega = A^{-1} \int d \bb r\,\tv{\omega_z^2(\bb r,t)}$ in
PT, where $\tv{F(t)}$ represents the time-average of $F(t)$ and
$A$ is the area of the system. We carry out our calculations in
Fourier space so the total number of modes $\hpsi(\kvec,t)$
considered is infinite. We apply a {\it Hartree-Fock
approximation} (HFA) to PT in which four-point correlation
functions are approximated by products of two-point correlation
functions. We also assume the time-averaged two-point correlation
function $\hat{C}_\psi(\kvec_1,\kvec_2) \equiv \tv{\hpsi(\kvec_1,t)
\hpsi(\kvec_2,t)}$ obeys a Gaussian distribution, i.e.,
$\hat{C}_{\psi}(\kvec_1,\kvec_2) = \hj(\kvec_1) \exp[-\hf (\kvec_1 + \kvec_2)^2
\xi_\psi^2]$, where $\xi_{\psi}$ is a length determining the
correlation between different modes.
 This approximation
is based on the observation that the patterns in PT are
disordered
on long length scales and quasi-ordered on short length scales.  This
suggests that the correlation between different wavevectors ${\vec k}_1$ and
${\vec k}_2$ is small unless these vectors are close to each other.  This
approximation seems physically intuitive for PT, in which there is a
random superposition of rolls of different orientations.  Given the above
assumptions, we derive
$J$ and $\Omega$ in
terms  of the time-averaged and azimuthally averaged structure factor
$S(k) \equiv \tv{\hpsi^*(\kvec,t) \hpsi(\kvec,t)}/J$.
We further assume that $S(k)$ obeys a scaling form
$k S(k) = \xi_2 {\mathcal F}[(k - k_{max}) \xi_2]$, in which $\xi_2$ is the two-point
correlation length, ${\mathcal F}(x)$ is the scaling function and $k_{max}$ is the
peak position of $k S(k)$. Applying this assumption,
we obtain explicit formulas for both $J$ and $\Omega$. More precisely, we
find
that $J = J_0 \epsilon - J_\xi \xi_2^{-2}$,
which depends on
unknown but experimentally measurable parameters, where
$\epsilon = (R - R_c)/R_c$ is the
reduced control parameter and $J_0$ and $J_\xi$ are both known.
We also find that $\Omega = \omega J^2/\xi_2$
where $\omega$ is related to the
width of the scaling function
${\mathcal F}(x)$.
Furthermore, by assuming power law behaviors such that $J \sim
\epsilon^\mu$,
$\xi_2 \sim \epsilon^{-\nu}$ and $\Omega \sim \epsilon^\lambda$, we predict
from our theory that
$\mu = 1$, $\nu \ge 1/2$ and $\lambda = 2 \mu + \nu$
for PT. This prediction and the scaling assumption for
$S(k)$ have been verified by
our numerical solutions.

Our paper is organized as follows. In Sec. \srf{basic},
we introduce the GSH model in Fourier space and derive the basic
formulas governing the time-averaged convective current $J$ and
the time-averaged mean square vorticity  $\Omega$ for any pattern in RBC.
In Sec. \srf{rpa} we introduce the HFA of
our theory and use it to calculate explicitly $J$ and $\Omega$
for PT. The results are expressed in terms of the structure
factor $S(k)$.
Sec. \srf{scaling} includes two parts. We first postulate
the scaling form of the structure factor $S(k)$ and expand both
$J$ and $\Omega$
in the leading order of $1/\xi_2$.
We then define power law behaviors for $J$, $\xi_2$
and $\Omega$ in PT and compare the results from
our theory and our numerical work \cite{xi_li_97,xi_li_00}, 
which agree very well
for both the exponents and the amplitudes.
In the last section, we summarize our results and
discuss some related issues.

\section{Basic formulas} \lbl{basic}

In the GSH model, the order parameter $\psi(\bb r,t)$
satisfies \cite{sw_ho_77,cr_80,si_zi_81,li00} \bge
        \tau_0\left[\prtt \psi + \bb U \cdot \bb \nabla \psi\right]
        = \left[\epsilon -(\xi_0^2/4 k_c^2)({\bb \nabla}^2
          + k_c^2)^2 \right] \psi
               - N[\psi].
                                \lbl{gsh}
\nde
Here $N[\psi]$ is the nonlinear term to be specified soon and
$\bb U(\bb r)$ is the mean flow velocity given by
$\bb U(\bb r) = \vna \zeta(\bb r,t) \times {\bb e}_z$,  in which
\bge
        \left[\prtt -\sigma \lpl \right] \lpl \zeta
                = g_m {\bb e}_z \cdot \left[ \vna(\lpl \psi)
                \times \vna \psi\right]. \lbl{mf}
\nde
In the GSH equations, the reduced Rayleigh number $\epsilon \equiv (R/R_c)
-1$
is the control parameter, in which  $R$ and $R_c$ are
the Rayleigh number and its critical value at onset.
The Prandtl number $\sigma$ parameterizes the fluid.
While $k_c$ is the critical wavenumber at onset, other parameters
model the properties of the system. More precisely, one takes
\cite{cr_80,para}
$R_c = 27 \pi^4/4$, $k_c = \pi/\sqrt{2}$,
$\tau_0 = 2 (1 + \sigma^{-1})/3 \pi^2$,
$\xi_0^2 = 8/3 \pi^2$, and
$g_m = 6$.

It is easier to analyze the GSH equations theoretically
in Fourier space than in real space.
By convention, we define the Fourier transformation and
its inverse transformation of an
arbitrary function $F(\bb r)$ as
$\hat{F}(\kvec) = \frac{1}{A} \int d \bb r \, e^{- i \kvec \cdot \bb r} F(\bb r)$
and $F(\bb r) = \sum_{\bb k} \hat{F}(\kvec) e^{i \kvec \cdot \bb r}$,
where $A$ is the area of the system.
Note that $\hat{F}^*(\kvec) = \hat{F}(-\kvec)$ for any real function $F(\bb r)$.
It is easy to check that Eq. \rf{gsh} can be rewritten in Fourier space as
\bge
\tau_0\prtt \hat{\psi}(\kvec)
 + \hat{V}(\kvec)
 = r(\epsilon; k)
  \hat{\psi}(\kvec) -\hat{N}(\kvec),
  \lbl{ftgsh}
\nde
where $V(\bb r) = \tau_0 \bb U \cdot \vna \psi$ and
\bge
r(\epsilon; k) = \epsilon - \xi_0^2 (k^2 - k_c^2)^2/4 k^2_c.
\lbl{r}
\nde
The nonlinear $\hat{N}(\kvec)$ term has been  evaluated at onset \cite{cr_80},
\bge
\hat{N}(\kvec) = \sum_{\kvec_2, \kvec_3}
g(\hat{k} \cdot \hat{k}_2)
\hat{\psi}^*(\kvec_2) \hpsi(\kvec_3) \hpsi(\kvec+\kvec_2-\kvec_3),
\lbl{nl}
\nde
where the coupling constant $g(\cos \alpha)$ is
given in Ref. \cite{cr_80} with $\alpha$ the angle between $\kvec$ and
$\kvec_2$. Rigorously speaking, the exact forms of Eqs. \rf{gsh}, \rf{mf} and
\rf{nl} are derived near onset and deviations from them in real physical
systems are possible for large enough $\epsilon$.
But we disregard such complexity and take them
as our model for further study.

One may take an adiabatic approximation $(\partial_t = 0)$ in Eq. \rf{mf}
by neglecting the first term on the left-hand side.
This term is small in comparison with the other terms, which
can be verified by applying the same perturbation as that in
phase dynamics \cite{po_ma_79}. With this approximation,
it now is easy to solve Eq. \rf{mf} for $\hat{\zeta}(\kvec)$,
which indicates that the mean-flow field is slaved by the $\psi(\bb r,t)$
field. We are also interested in the
vertical vorticity $\omega_z(\bb r) = - \lpl \zeta(\bb r)$.
From Eq. \rf{mf}, it is straightforward to get that
\bge
\hat{\omega}_z(\kvec) = k^2 \hat{\zeta}(\kvec)
= \sum_{\kvec_2} f(\kvec; \kvec_2) \hpsi(\kvec_2) \hpsi(\kvec - \kvec_2),
\lbl{vtct}
\nde
where, with an exchange of index $\kvec_2 \to \kvec - \kvec_2$,
\bge
f(\kvec; \kvec_2) = \frac{g_m}{2 \sigma k^2} (k^2 - 2 \kvec \cdot \kvec_2)
({\bb e}_z \cdot \kvec_2 \times \kvec). \lbl{cplf}
\nde
Applying these results, one may easily
evaluate the mean-flow contribution to Eq. \rf{ftgsh}, which is given by
\bge
\hat{V}(\kvec) = \sum_{\kvec_2, \kvec_3}   v(\kvec; \kvec_2; \kvec_3)
\hat{\psi}^*(\kvec_2) \hpsi(\kvec_3) \hpsi(\kvec + \kvec_2 -\kvec_3),
\lbl{ftmf}
\nde
where
\bge
v(\kvec; \kvec_2; \kvec_3) = \frac{g_m \tau_0}{2 \sigma}
\frac{[{\bb e}_z \cdot \kvec \times (\kvec_3 - \kvec_2)]
  [{\bb e}_z \cdot \kvec_3 \times \kvec_2] (k_2^2 -k_3^2)}
{|\kvec_3 -\kvec_2|^4} \, .
\lbl{cplv}
\nde
Notice that the coupling constant $v(\kvec;\kvec_2;\kvec_3)$ is zero under two
conditions: (1) If all $\kvec$ allowed in $\hpsi(\kvec)$ point at one single
direction, say $\hat{k}$;
or, (2) if all $\kvec$ lie on one single ring, say $|\kvec| = k$. For this reason,
ordered states  such as parallel rolls, hexagons,
concentric rings, etc., do not have significant mean-flow couplings.
Furthermore, the coupling constant $v(\kvec;\kvec_2;\kvec_3)$
seems to have a pole at $\kvec_2 = \kvec_3$. Assume that
$\kvec_3 = \kvec_2 + \q$ with $\q$ very small; then
$v \sim ({\bb e}_z \cdot \kvec \times \q)
({\bb e}_z \cdot \kvec_2 \times \q)(\kvec_2 \cdot \q)
/q^4$. A pole normally exists (since $v \sim 1/q$)
unless $\q \| \kvec$, or $\q \| \kvec_2$, or $\q \perp \kvec_2$.

In this paper, we will mainly focus on two global quantities: One
is the total convective current defined by
\bge
J(t) = \frac{1}{A} \int d \bb r \, \psi^2(\bb r, t)
= \sum_{\kvec} \hat{J}(\kvec,t) \quad {\rm with} \quad
\hat{J}(\kvec,t) = \hpsi^*(\kvec,t) \hpsi(\kvec,t);
\lbl{cvcrnt}
\nde
the other is the mean square vorticity defined by
\bgea
\Omega(t) = \frac{1}{A} \int d \bb r \, \omega_z^2(\bb r, t)
= && \sum_{\kvec_1,\kvec_2,\kvec_3,\kvec_4}   f(\kvec_1+\kvec_2;\kvec_2)
f(\kvec_1+\kvec_2;\kvec_3)  \nonumber \\
 &&\times \hpsi^*(\kvec_1) \hpsi^*(\kvec_2) \hpsi(\kvec_3) \hpsi(\kvec_4)
\delta_{\kvec_1+\kvec_2,\kvec_3+\kvec_4}.
 \lbl{vtctcrnt}
\ndea
Notice that
$f(\kvec_1+\kvec_2;\kvec_2) \sim (k_1^2 -k_2^2)({\bb e}_z \cdot \kvec_2 \times \kvec_1)$.
So $\Omega(t) =0$
if all the wavenumbers allowed in $\hpsi(\kvec)$
point at one single direction $\pm \hat{k}$
or lie on one single ring $|\kvec| = k$.
In other words, the vorticity must be generated by couplings
between modes of different $k$ and $\hat{k}$.
From Eq. \rf{ftgsh}, it is easy to derive that
\bgea
\tau_0 \prtt \hj(\kvec,t) = && 2 r(\epsilon; k) \hj(\kvec,t)
-  \sum_{\kvec_2,\kvec_3} 
\left[g(\hat{k} \cdot \hat{k}_2)+v(\kvec; \kvec_2; \kvec_3)\right]
\nonumber \\ & \times &
\left[\hpsi^*(\kvec) \hpsi^*(\kvec_2) \hpsi(\kvec_3) \hpsi(\kvec+\kvec_2-\kvec_3)
  + \mathop{\rm c.c.}\nolimits \right].
\lbl{keqcvcrnt}
\ndea
In principle, this is the equation determining the structure of the
convective current $\hj(\kvec,t)$ which, however,
is beyond our present goal.
Now applying the relations $\hpsi^*(\kvec) = \hpsi(-\kvec)$ and
$v(\kvec; \kvec_2; \kvec_3)=-v(\kvec; \kvec_3;\kvec_2)$, and,
exchanging the summation indices $\kvec \to -\kvec$, $\kvec_{2,3} \to - \kvec_{2,3}$
for the
$g$ terms and  $\kvec \to \kvec + \kvec_2 -\kvec_3$, $\kvec_2 \leftrightarrow \kvec_3$ for
the $v$ terms, one obtains from the above equation and the definition of
$J(t)$ that
\bgea
\hf \tau_0 \prtt J(t) = && \sum_{\kvec} r(\epsilon; k) \hj(\kvec,t)
\nonumber \\
&-& \sum_{\kvec_1, \kvec_2,\kvec_3} g(\hat{k}_1 \cdot \hat{k}_2)
\hpsi^*(\kvec_1) \hpsi^*(\kvec_2) \hpsi(\kvec_3) \hpsi(\kvec_1+\kvec_2-\kvec_3).
\lbl{eqcvcrnt}
\ndea
This is the equation determining the total convective
current $J(t)$.
Notice that the $v$ terms vanish from this equation, which
can also be derived directly from Eq. \rf{gsh}
by converting the corresponding integral in Eq. \rf{cvcrnt} into
a surface term. In general, the $v$ couplings affect
$J(t)$ implicitly by modifying  its structure $\hj(\kvec,t)$
unless, of course,  $v \equiv 0$.

For stationary states, the convective current and the mean square vorticity
 are time-independent. This, however, is no long true if the state
is spatiotemporal chaotic.  For a spatiotemporal chaotic state, these
two quantities  normally fluctuate in time around some well-defined averaged
values: see Refs. \cite{xi_li_97,xi_gu_95}.
While the fluctuations appear chaotic in time, they are
relatively small in comparison with their averaged values. For simplicity,
we
only consider the two corresponding time-averaged quantities in our theory.
We now introduce the time-average operator $\mathcal T$ defined by
${\mathcal T} F(t) \equiv \tv {F(t)}
= \lim_{T \to +\infty} \frac{1}{T} \int^T_0 dt F(t)$.
Applying $\mathcal T$ to Eq. \rf{eqcvcrnt} yields
\bgea
&& \sum_{\kvec} r(\epsilon; k) \tv{\hj(\kvec,t)} \nonumber \\
&& - \sum_{\kvec_1, \kvec_2,\kvec_3,\kvec_4} g(\hat{\kvec}_1 \cdot \hat{\kvec}_2)
\tv{\hpsi^*(\kvec_1) \hpsi^*(\kvec_2) \hpsi(\kvec_3) \hpsi(\kvec_4)} \,
\delta_{\kvec_1+\kvec_2,\kvec_3+\kvec_4} = 0.
\lbl{tveqcvcrnt}
\ndea
In the next section, we show how, under a certain assumption,
to calculate the time-averaged convective current of PT
from this equation. The time-averaged mean square vorticity  can be obtained
with $\mathcal T$ acting on Eq. \rf{vtctcrnt}. For simplicity,
we denote from now on
$\hj(\kvec) = \tv{\hj(\kvec,t)}$, $J = \tv{J(t)}$ and $\Omega = \tv{\Omega(t)}$.

Finally we introduce the time-averaged structure factor defined
by  $\hat{S}(\kvec) = \hj(\kvec)/J$ with
$\sum_{\kvec} \hat{S}(\kvec) = 1$,
and the corresponding averages
$\pv{\hat{F}}_\kvec
= \sum_{\kvec} \hat{S}(\kvec) \hat{F}(\kvec)$.
With this notation, the first term in Eq. \rf{tveqcvcrnt} can be rewritten
as
$\pv {r(\epsilon)}_\kvec J$. If the $k$-dependence and the angular dependence
in $\hat{S}(\kvec)$ can be separated, then it is more convenient to define
$\hat{S}(\kvec) = (2 \pi)^2 A^{-1} S(k) \Phi(\alpha)$ with
$\int^\infty_0 d k \, k S(k) = 1$ and
$\int^{2 \pi}_0 d \alpha \, \Phi(\alpha) = 1$,
where $\alpha$ is the angle between $\kvec$ and some reference direction.
Here
the discrete $\kvec$ lattice has been converted
into a continuous one. So a proper phase factor has been taken into account.
Notice also that $\Phi(\pi + \alpha) = \Phi(\alpha)$ since
$\hat{S}(-\kvec) = \hat{S}(\kvec)$.
Now the corresponding averages with respect to $S(k)$ and $\Phi(\alpha)$
are defined as
$\pv F_k = \int^\infty_0 d k \, k S(k) F(k)$ and
$\pv F_\alpha = \int^{2 \pi}_0 d \alpha \, \Phi(\alpha) F(\alpha)$.

\section{Hartree-Fock Approximation for PT} \lbl{rpa}

\subsection{Convective Current} \lbl{rpa_cc}

In order to calculate the time-averaged convective current and
the time-averaged mean square vorticity from
Eqs. \rf{tveqcvcrnt} and \rf{vtctcrnt},
it is obvious that more information about the corresponding
state is needed. For this purpose,
we notice that the instantaneous patterns in PT
are irregular and random in space
\cite{xi_li_97}.
So, if we write
$\hpsi(\kvec,t) = \hat{\rho}(\kvec,t) e^{i \hat{\phi}(\kvec,t)}$ with both
$\hat{\rho}(\kvec,t)$ and $\hat{\phi}(\kvec,t)$ real, the phase
$\hat{\phi}(\kvec,t)$ seems to have a rather complicated, irregular
time-dependence. This suggests that two modes of different $\kvec$'s
are poorly time-correlated. Because of this, we think it is
justified to adapt a {\it Hartree-Fock Approximation} (HFA) to PT
in which
a four-point correlation function is approximated by products
of  two-point correlation functions, i.e.,
\bgea
&&\hat{C}_{\psi}^{(4)}(\kvec_1,\kvec_2,\kvec_3,\kvec_4) \equiv
\tv{\hpsi(\kvec_1,t) \hpsi(\kvec_2,t)\hpsi(\kvec_3,t)\hpsi(\kvec_4,t)} \nonumber \\
&&\simeq \hat{C}_{\psi}(\kvec_1,\kvec_2) \hat{C}_{\psi}(\kvec_3,\kvec_4)
+ \hat{C}_{\psi}(\kvec_1,\kvec_3) \hat{C}_{\psi}(\kvec_2,\kvec_4)
+ \hat{C}_{\psi}(\kvec_1,\kvec_4) \hat{C}_{\psi}(\kvec_2,\kvec_3),
\lbl{rpa_four}
\ndea
where we denote
\bge
\hat{C}_{\psi}(\kvec_1,\kvec_2) \equiv \tv{\hpsi(\kvec_1,t)\hpsi(\kvec_2,t)}.
\lbl{tw_pt}
\nde
Apparently $\hat{C}_{\psi}(\kvec,-\kvec) = \hj(\kvec)$.

Now it is essential to know the behavior of $\hat{C}_{\psi}(\kvec_1,\kvec_2)$
in PT.  We notice that
the patterns in PT are disordered at large scales
but quasi-ordered inside some smaller domains.
From these, we speculate that the correlation between two modes of
different $\kvec$'s are poor if their $\kvec$'s are quite different
but good if their $\kvec$'s are very close to each other. Hence, we
postulate a Gaussian form for $\hat{C}_{\psi}(\kvec_1,\kvec_2)$ in PT, i.e.,
\bge
\hat{C}_{\psi}(\kvec_1,\kvec_2)
= \hj(\kvec_1) \exp[-\hf (\kvec_1 + \kvec_2)^2 \xi_\psi^2], \lbl{c_2}
\nde
where $\xi_\psi$ is a length determining the correlation between different
modes.  We expect $\xi_\psi$ to be large.

We now calculate the convective current $J$ from
Eqs. \rf{tveqcvcrnt}, \rf{rpa_four} and \rf{c_2}. One
may rewrite Eq. \rf{tveqcvcrnt} in a continuous Fourier space as
\bgea
\pv{r(\epsilon,k)}_\kvec J
=\frac{A^4}{(2 \pi)^8} &&\int d \kvec_1 d \kvec_2 d \kvec_3 d \kvec_4 \,
\frac{(2 \pi)^2}{A} \delta(\kvec_1 + \kvec_2 - \kvec_3 -\kvec_4)\,
g(\hat{k}_1 \cdot \hat{k}_2) \nonumber \\
&&\times
\hat{C}_{\psi}^{(4)}(-\kvec_1,-\kvec_2,\kvec_3,\kvec_4).
\lbl{tveqcvcrnt_rpa}
\ndea
Inserting Eqs. \rf{rpa_four} and \rf{c_2} into this equation,
one finds after some algebra that
\bge
J = \frac{4 \pi \xi_\psi^2}{A} \frac{\pv{r(\epsilon,k)}_k}{g_{PT}},
\lbl{tvcvcrnt_rpa2}
\nde
in addition to the conduction solution $J = 0$.
Here $\hat{S}(\kvec)$ is assumed to be azimuthally uniform, and
$g_{PT}$ is defined as
\bge
g_{PT} = g(-1) + \frac{2}{\pi} \int^\pi_0 d \alpha\,g(\cos\alpha).
\lbl{g_inf}
\nde
Using the explicit formula given in Ref. \cite{cr_80} for free-free
boundaries, one has that
\bge
g_{PT} = 0.855922 + 0.0458144 \sigma^{-1} + 0.0709326 \sigma^{-2},
\lbl{g_pt}
\nde
where $\sigma$ is the Prandtl number.

If one takes
the limit $\xi_\psi \to + \infty$, the two-point correlation function
$\hat{C}_{\psi}(\kvec_1,\kvec_2)$ in Eq. \rf{c_2} reduces to
$(A/2 \pi \xi_\psi^2) \hj(\kvec_1) \delta_{\kvec_1,-\kvec_2}$. This,
by applying the inverse Fourier transformation,
leads to the translation invariance of the
the two-point correlation function
$C(\bb r_1,\bb r_2) \equiv \tv{\psi(\bb r_1,t) \psi(\bb r_2,t)}/J$, i.e.,
$C(\bb r_1,\bb r_2) = C(\bb r_1 - \bb r_2)$.
Under the same limit,
one also has $\xi_\psi = [A/2 \pi]^{1/2}$
since $\hat{C}_{\psi}(\kvec,-\kvec) = \hj(\kvec)$. In this case, one has
\bge
J = \frac{2 \pv{r(\epsilon,k)}_k}{g_{PT}}.
\lbl{tvcvcrnt_rpa}
\nde
From now on, we simply choose $\xi_\psi = [A/2 \pi]^{1/2}$.

\subsection{Mean Square Vorticity} \lbl{rpa_vc}

We now use the HFA to calculate the time-averaged mean square vorticity.
From Eq. \rf{vtctcrnt}, one finds that
\bgea
\Omega
=\frac{A^4}{(2 \pi)^8} &&\int d \kvec_1 d \kvec_2 d \kvec_3 d \kvec_4 \,
\frac{(2 \pi)^2}{A} \delta(\kvec_1 + \kvec_2 - \kvec_3 -\kvec_4)
f(\kvec_1+\kvec_2;\kvec_2) f(\kvec_1+\kvec_2;\kvec_3)
\nonumber \\
&&\times
\hat{C}_{\psi^4}(-\kvec_1,-\kvec_2,\kvec_3,\kvec_4),
\lbl{eq_tvvtctcrnt}
\ndea
where, from Eq. \rf{cplf},
\bgea
&&f(\kvec_1+\kvec_2;\kvec_2) f(\kvec_1+\kvec_2;\kvec_3) \delta(\kvec_1+\kvec_2-\kvec_3-\kvec_4)
\nonumber \\
&&= \frac{g_m^2}{4 \sigma^2} \frac{(k_1^2 - k_2^2) (k_3^2 - k_4^2)
  ({\bb e}_z \cdot \kvec_1 \times \kvec_2)
  ({\bb e}_z \cdot \kvec_3 \times \kvec_4)}
{|\kvec_1 + \kvec_2|^2\, |\kvec_3 + \kvec_4|^2}
\delta(\kvec_1+\kvec_2-\kvec_3-\kvec_4). \lbl{eq_ff}
\ndea
Notice that there is a singular point $\kvec_1 + \kvec_2 = 0$
in the above expression:
its value depends on how the point is approached.

Following the same calculation as for $J$ and
assuming that $\hat{S}(\kvec)$ is azimuthally uniform,
one finds after some algebra that
\bge
\Omega = \frac{g_m^2 J^2}{4 \sigma^2}
\int dk_1\,k_1 S(k_1) \,\int d k_2 \, k_2 S(k_2) \,\Delta(k_1;k_2),
\lbl{tvvtctcrnt}
\nde
where  we have
chosen $\xi_\psi = [A/2 \pi]^{1/2}$, and
\bge
\Delta(k_1;k_2) = \frac{1}{4} |k_1^2 - k_2^2|\,
[k_1^2 + k_2^2 - |k_1^2 - k_2^2|], \lbl{Delta_fr}
\nde
which has a second-order singularity at $k_1 = k_2$ and is due to the
singularity in Eq. \rf{eq_ff}.

\section{Scaling Relations in PT} \lbl{scaling}

\subsection{General} \lbl{scaling_gen}

To evaluate the convective current $J$ and the mean square vorticity,
$\Omega$,
one must know the structure factor $S(k)$ which, however, is beyond our
present theory. We thus  turn to phenomenological arguments.
We define a two-point correlation length as
\bge
\xi_2 = \left[\pv{k^2}_k - \pv{k}_k^2\right]^{-1/2} \lbl{lngth}.
\nde
Then we
assume that the structure factor satisfies the following {\it scaling} form
\bge
k S(k) = \xi_{2} {\cal F}[(k - k_{max}) \xi_2],  \lbl{s_scaling}
\nde
where $k_{max}$ is the peak position
of $k S(k)$ and  ${\cal F}(x)$ is the scaling function satisfying
$\int_{-\infty}^\infty d x\, {\cal F}(x) = 1$.
[Since $k \ge 0$ in $k S(k)$,
the lower limit for ${\cal F}(x)$ is $- k_{max} \xi_2$, which we approximate
by $-\infty$.]
Inserting $k = k_{max} + x \xi_2^{-1}$
and Eq. \rf{s_scaling} into Eq. \rf{lngth},
one gets that $\pv{x^2}_x - \pv{x}_x^2 = 1$, where we have used the notation
$\pv{F(x)}_{x}=\int_{-\infty}^\infty d x \, {\cal F}(x) F(x)$.
It is also easy to see that $\pv{k}_k = k_{max} + \xi_2^{-1} \pv{x}_x$.

For very large $\xi_2$, one may take $k = k_{max} + x \xi_2^{-1}$ in
Eqs. \rf{tvcvcrnt_rpa} and \rf{tvvtctcrnt} and
expand the results in order of $1/\xi_2$.
It is easy to find from Eqs. \rf{r} and \rf{tvcvcrnt_rpa},
and from $k_{max} = k_c + O(\epsilon)$ that
\bge
J = \frac{2}{g_{PT}}
\left[\epsilon - \pv{x^2}_x\frac{\xi_0^2}{\xi_2^2}\right]
+ O\left(1/\xi_2^3,\epsilon/\xi_2,\epsilon^2\right),
\lbl{tvcvcrnt_rpa_asmp}
\nde
for small enough $\epsilon$ and large enough $\xi_2$.
This expression depends on two unknown but experimentally measurable
quantities $\xi_2$ and $\pv{x^2}_x = 1 + \pv{x}_x^2$. If ${\mathcal F}(x)$ is
symmetric, then $\pv{x}_x = 0$ and $\pv{x^2}_x = 1$. In general, however,
one
has $\pv{x^2}_x \ge 1$.
Notice also that although
mean-flow couplings are not explicitly present in Eq. \rf{tveqcvcrnt}, they
affect the value of the convective current via the structure
factor $\hat{S}(\kvec)$ [see Eq. \rf{keqcvcrnt}]
and the two-point correlation  length $\xi_2$.
For PT, since
$\xi_2 \simeq \frac{3}{2} \xi_0/\epsilon^\hf$ \cite{xi_li_97}, the
contribution
from $\xi_2$ reduces the value of $J$ quite significantly.
This feature is absent in the
simple theory presented earlier \cite{xi_li_00}.

One may evaluate the mean square vorticity in the same way. From
Eqs. \rf{tvvtctcrnt} and \rf{Delta_fr}, one gets for free-free boundaries
that
\bge
\Omega \approx
\frac{g_m^2 k_c^3}{4 \sigma^2}\pv{|x_1-x_2|}_{x_1,x_2}\frac{J^2}{\xi_2},
\lbl{Omega_scaling_fr_fr}
\nde
where we have used $k_{max} \approx k_c$.
The quantity $\pv{|x_1-x_2|}_{x_1,x_2}$ is related to the width of the
scaling
function ${\mathcal F}(x)$ and, since
$\pv{|x_1-x_2|}_{x_1,x_2} \le \sqrt{\pv{(x_1-x_2)^2}_{x_1,x_2}} = \sqrt{2}$,
we expect $\pv{|x_1-x_2|}_{x_1,x_2}$ to be of order of unity.
Clearly, by
phenomenological arguments,
we can express $J$ and $\Omega$ in PT in terms of measurable
quantities.

\subsection{Power Laws} \lbl{scaling_pt}

For PT, we further assume  power law behaviors for the two-point correlation
length, the convective current and the mean square vorticity
such  as
\bge
\xi_2 \approx \xi_{2,0} \epsilon^{-\nu}, \quad
J \approx J_0 \epsilon^\mu \quad {\rm and} \quad
\Omega \approx  \Omega_0 \epsilon^\lambda. \lbl{scaling_exponts}
\nde
Then, from Eq. \rf{Omega_scaling_fr_fr}, we find
the following scaling relation
\bge
\lambda = 2 \mu + \nu.
\lbl{scaling_fr_fr}
\nde
Recalling Eq. \rf{tvcvcrnt_rpa_asmp},
one obtains that
\bge
J \approx \frac{2}{g_{PT}}
\left[\epsilon -\pv{x^2}_x
  \frac{\xi_0^2}{\xi_{2,0}^2} \epsilon^{2 \nu}\right]
\approx J_0 \epsilon^{\mu}. \lbl{J_scaling}
\nde
Since $J$ is positive by definition, the values of the
exponents  satisfy
\bge
\mu = 1, \quad \nu \ge 1/2 \quad
{\rm and} \quad \lambda = 2 + \nu \ge 5/2.
\lbl{exponents_fr}
\nde
It is very likely that $\nu = 1/2$, hence, $\lambda = 5/2$. If so,
then one finds from Eqs. \rf{tvcvcrnt_rpa_asmp} and \rf{Omega_scaling_fr_fr}
that
\bge
J_0 = \frac{2}{g_{PT}}
\left[1 - \pv{x^2}_x \frac{\xi_0^2}{\xi_{2,0}^2}\right]
\quad {\rm and} \quad
\Omega_0 =
\frac{g_m^2 k_c^3 J_0^2}{4 \sigma^2 \xi_{2,0}}
\pv{|x_1-x_2|}_{x_1,x_2},
\lbl{amps_fr_fr}
\nde
which depend on three phenomenological parameters $\xi_{2,0}$, $\pv{x^2}_x$
and $\pv{|x_1-x_2|}_{x_1,x_2}$. If $\nu > 1/2$, then $J_0 = 2/g_{PT}$
since the $\epsilon^{2 \nu}$ term
in Eq. \rf{J_scaling} contributes only to the leading correction to scaling.
It is interesting to notice that the amplitude equations coupled
with mean-flow \cite{si_zi_81} predicts for free-free boundaries
that $\Omega \sim \epsilon^{5/2}$ for almost perfect parallel rolls.

We now verify our predictions for the power laws in PT
by our numerical solutions. We have carried out
large-scale numerical calculations of the three-dimensional Boussinesq
equations  under free-free boundaries for fluids of $\sigma = 0.5$
\cite{xi_li_97,xi_li_00}. We have confirmed in Ref. \cite{xi_li_97}
that the structure factor in PT satisfies
the scaling form \rf{s_scaling}.
From Table \ref{table_pt}, one can see that our theoretical
and our numerical results
are in very good agreement for the exponents. The scaling relation
Eq. \rf{scaling_fr_fr} is confirmed within our numerical uncertainties.
The comparison between the corresponding amplitudes, however, is only
moderately successful. Calculations of $\xi_{2,0}$, $\pv{x^2}_x$ and
$\pv{|x_1 - x_2|}_{x_1,x_2}$
are obviously beyond the present theory, so we take
our numerical result for $\xi_{2,0}$.
Since our numerical results for $\pv{x^2}_x$ and $\pv{|x_1 -
x_2|}_{x_1,x_2}$
are too sensitive to the large value cutoff to be meaningful,
we assume equalities in
$\pv{x^2}_x = 1 + \pv{x}_x^2 \ge 1$ and
$\pv{|x_1 - x_2|}_{x_1,x_2} \le \sqrt{\pv{(x_1-x_2)^2}_{x_1,x_2}} =
\sqrt{2}$.
 From Eqs. \rf{amps_fr_fr} and \rf{g_pt}, one gets
$J_0 \simeq 0.972$, which is about $20 \%$ larger than the numerical value.
A non-zero value of $\pv{x}_x$ will apparently reduce the theoretical
value of $J_0$ in the right direction.
On the other hand, one finds that
$\Omega_0 \simeq 454.7 \pv{|x_1-x_2|}_{x_1,x_2} = 643.0$ as an upper bound,
which is about ten times larger
than our numerical result. Nevertheless,
we note that while our theory is based on the
two-dimensional GSH equations, our numerical calculations are done for
the three-dimensional Boussinesq equations. Although the former is very good
in reproducing qualitative features of RBC, it
may not be quantitatively accurate in modeling RBC \cite{cr_ho_93}.
So one should be cautious in comparing
the results from the GSH equations
with those from real experiments
or those from numerical calculations with hydrodynamical equations.

\section{Discussion} \lbl{disc}

Our phenomenological theory for PT in RBC
depends on two basic assumptions. In Sec. \srf{rpa},
we adapt a Hartree-Fock approximation to PT and postulate
a Gaussian form for the time-averaged
two-point correlation function.
In Sec. \srf{scaling}, we further assume that
the structure factor
satisfies a scaling form such as
$k S(k) = \xi_2 {\mathcal F}[(k- k_{max})\xi_2]$. In comparison with
similar scaling forms in critical phenomena, critical dynamics and
phase ordering, we find it necessary to replace $k$ with
$k - k_{max}$ in the scaling form. The physical origin of
this replacement is due to  the fact that patterns in RBC have an intrinsic
wavenumber, which is close to $k_c$. By the same reason, we find it
necessary
to seek the scaling form of $k S(k)$ instead of $S(k)$, where the $k$
factor comes from $d \kvec = k\, dk\, d \alpha$ in two-dimensional $k$-space.
The physical
origin of this scaling invariance in PT is yet unknown.
In Sec. \srf{scaling}, we have confirmed the scaling form
of $S(k)$ within our numerical accuracy.
It is not clear
currently whether the scaling form breaks down for small or large $k$.

In summary, we present a phenomenological theory for PT in RBC. We calculate
analytically the time-averaged convective current $J$ and the time-averaged
mean square vorticity  $\Omega$
as functions of $\epsilon$ and $\xi_2$.
We believe that our theoretical results
will be useful in understanding the complicated behavior of STC in RBC.
We also believe that our theory provides a new
approach to STC and
also raises some interesting questions.
For example, how can one
calculate the structure factor $S(k)$ and the two-point
correlation length $\xi_2$ analytically? Is it possible that certain global
quantities in STC form a complete set in the same way as temperature,
pressure and density do for thermodynamic systems?  Can we derive
some effective variational principle in terms of global quantities?
How far can we apply the ideas in critical phenomena to study STC?
Since our assumptions are quite general, it will also be interesting to see
whether our theory can be generalized to STC in other systems
\cite{cr_ho_93}.

\begin{center} Acknowledgment \end{center}

X.J.L and J.D.G are supported by the National Science Foundation
under Grant No. DMR-9596202. H.W.X. is supported by Research Corporation
under Grant No. CC4250.
Numerical work reported
here are carried out on the Cray-C90 at the Pittsburgh
Supercomputing  Center and Cray-YMP8 at the Ohio Supercomputer Center.
X.J.L. also acknowledges the support of NSF under Grant No. DMR-9876864
when this paper was revised.

%

\begin{table}
\caption{Time-averaged convective current $J \approx J_0 \epsilon^\mu$,
time-averaged mean square vorticity $\Omega \approx \Omega_0
\epsilon^\lambda$
and two-point correlation length $\xi_2 \approx \xi_{2,0} \epsilon^{-\nu}$
in PT with $\sigma = 0.5$. For theoretical result of $\nu$,
we assume equality in Eq. (34). See also discussions in Sec. IV(B).}
\lbl{table_pt}

\begin{tabular}{lcccccc}
 & $\mu$ & $\nu$ & $\lambda$ & $\xi_{2,0}$ & $J_0$ & $\Omega_0$ \\ \hline
Numerics & $1.034 \pm 0.025$ & $0.472 \pm 0.016$ & $2.55 \pm 0.10$
& $0.82 \pm 0.04$ & $0.787 \pm 0.019$ & $70.1 \pm 1.0$ \\
Theory & $1$ & $1/2$ & $5/2$ & ---
& $0.972$ & $643.0$ \\
\end{tabular}
\end{table}

\end{document}